\documentclass[aps,pra,showpacs,floatfix,twocolumn,superscriptaddress]{revtex4}
\usepackage{graphicx}
\usepackage{bm,amsmath,amssymb}
\usepackage{dcolumn}
\usepackage{mathrsfs}
\usepackage{cases}
\usepackage{color}
 \usepackage{ulem}
 \usepackage{indentfirst}

\newcommand{\be}{\begin{equation}}
\newcommand{\ee}{\end{equation}}
\newcommand{\bea}{\begin{eqnarray}}
\newcommand{\eea}{\end{eqnarray}}
\newcommand{\la}{\langle}
\newcommand{\ra}{\rangle}
\begin{document}
\title{Beats and broken-symmetry superfluid on a one dimensional anyon Hubbard model}
\author{Wanzhou Zhang}

\affiliation{College of Physics and Optoelectronics, Taiyuan University of Technology Shanxi 030024, China}
\author{Ernv Fan}
\affiliation{College of Physics and Optoelectronics, Taiyuan University of Technology Shanxi 030024, China}
\author{Tony C Scott}
\affiliation{College of Physics and Optoelectronics, Taiyuan University of Technology Shanxi 030024, China}
\author{Yunbo Zhang}
\affiliation{Institute of Theoretical Physics, Shanxi University, Taiyuan 030006, China }

\date{\today}
\begin{abstract}
By using the density matrix renormalization group and mean field methods,
the anyon Hubbard model is studied systematically on a one dimensional lattice.
The model can be expressed as a Bose-Hubbard model with a density-dependent-phase term.
When the phase angle is $\theta=0$ or $\theta=\pi$, the model will be equivalent to boson and pseudo fermion models, respectively.
 In the mean field frame, we find  a broken-symmetry superfluid (BSF), in which the $b^{\dagger}(b)$ operators on the nearest neighborhood sites have exactly opposite directions and   behave like a directed oscillation pattern.
By the density matrix reorganization group method, in the broken-symmetry superfluid, both the real and imaginary parts of  the correlation $b^{\dagger}_ib_{i+r}$ behave according to a {\it beat phenomenon} with $0<\theta<\pi$ in the form  $C_0e^{i k r}(-1)^{r}$
or behave like waves with different wavelengths in the form $C_0e^{i k r}$.
The distributions of the broken-symmetry superfluid phase and other phases are shown in the phase diagrams with different values
of $\theta$ and the direct phase transition between the two types of superfluid is observed.
The beats phenomenon is explained by double peaks of momentum distribution with two wave numbers ${k}_1$ and ${k}_2$ satisfying the condition $\frac{{k}_1-{k}_2}{{k}_1+{k}_2}<\frac{1}{3}$, which are expected to be observed in
the optical experiments.

\end{abstract}
\maketitle
\section{introduction}
Bosons and fermions,  are the two types of well-known elementary particles, respectively.
By exchanging the two bosons (fermions), the wave function  is  symmetric or  anti-symmetric, or updated
with a new phase factor $e^{i\theta}$, where $\theta=0$ for bosons, and $\theta=\pi$ for  pseudo fermions.
The exchange of two identical anyons will create a  phase angle
$ \theta$, which can be of any value. Anyons are governed by statistics which are intermediate between those of bosons and pseudo-fermions.  Anyons
have attracted much physical interest due to their novel
properties since the 1980s\cite{first}.
The anyon has become a very important concept in condensed matter
physics and Abelian anyons have  been detected  successfully and used in the understanding
of the fractional quantum Hall effect\cite{fqh}.

Experimentally, several schemes have been proposed
to search for the anyons in spin or boson models\cite{kitaev, longguilu,jianweipan, jingzhang, nmr} or
in cold atoms\cite{ex5,ex6,ex7,ex8,ex9}.
Theoretically,  through a  Jordan-Wigner transformation\cite{sta}, the anyon Hamiltonian can be mapped into
the Bose-Hubbard model with the tunneling terms coupled with a phase factor.
The picture of the Bose-Hubbard model is relatively clear making it easier to understand the effect of the
phase factor.

In the boson representation, there are have been many studies of anyons in the context of
multicomponent\cite{multicomponent}, entanglement\cite{entanglement}, dynamical\cite{dynamical},
ground-state\cite{ground1,ground2} and quantum walk\cite{quantumwalk} properties.
Ref.~\cite{sta}  studied the quantum phase transition of the
anyon Hubbard model, and found rich and interesting phases. Recently, Ref.~\cite{santos2} also proposed
an improved scheme to study the anyon Hubbard model and Ref.~\cite{tang} also studied the ground state
of the one dimensional  anyon model with open boundary conditions.

 The multiplication of
the phase $e^{i\theta}$ and tunneling amplitude $t$ varies from positive to minus signs, and
even to a complex number.
In spin language, effective ferromagnetic (non-frustrated) and anti-ferromagnetic (frustrated) tunneling emerges due to the modulation of $\theta$.
The frustrated tunneling will lead to a superfluid\cite{xfzhou2} condensed at different wave vectors, or a new supersolid without interactions\cite{santos}.
An interesting question arises: how does $e^{i\theta}$ affect the distribution and  transitions between the superfluid phases? The boson limit $\theta=0$ and pseudo  fermion limit $\theta=\pi$ are relatively clear, but in the range $0<\theta<\pi$, there
may be new phenomena, such as interesting momentum distributions\cite{tang}.

Herein, we study the anyon Hubbard model by both the mean field (MF) method and the density matrix renormalization group (DMRG) method\cite{dmrg1}.
In the MF frame,  we find a broken-symmetry SF (BSF) phase, in which the expectation value of the creation (annihilation) operator  $b^{\dagger}(b)$  behaves in a directed oscillation pattern.
By the DMRG method, the correlation $b^{\dagger}_ib_{i+r}$ behaves according to a beat phenomenon  with $0<\theta<\pi$
or to waves with different wavelengths.

The outline of this work is as follows.
Section \ref{sec:model} shows the Hamiltonian model, methods, and useful observables.
Section \ref{sec:result}  provides the MF results including the  BSF phase  and the phase diagrams.
A DMRG calculation is done in Sec.~\ref{sec:dmrg} and beats of the correlation are found and explained by the structure
of the double peak emerging in the momentum distributions.
Concluding comments are made in Sec.~\ref{sec:con}.

\section{The model, methods and observables}
\label{sec:model}
\subsection{model}
The starting point is the anyon-Hubbard Hamiltonian,
\be
H^a =-t\sum_{i=1}^{L}(a^{\dagger}_{i}a_{i+1}+h.c.)+\sum_{i}h_i\\
\label{ha}
\ee
where $a_i^{\dag} (a_i)$ is the anyon creation (annihilation) operator at site $i$,
$t$ is the single-anyon  hopping amplitude,
$L$ is the lattice size,
and $n_i=a_i^{\dag}a_i$ is the number operator of the anyons on site $i$.
In the term $h_i=\frac{U}{2} n_i    (n_i-1)-\mu n_i$, $U$ is the
on-site two-body interaction and $\mu$ is the chemical potential term.
By a Jordan-Wigner transformation\cite{sta},
\be
a_j=b_{j}e^{-i\theta\sum_{i=1}^{j-1}n_i},
\ee
where $b_i$ is the boson annihilation operator.
The anyon Hamiltonian $H^a$ can be re-expressed as a Bose-Hubbard model with a density dependent phase factor\cite{sta}:
\be
H^b =-t\sum_{i=1}^{L}(b^{\dagger}_{i}b_{i+1}e^{i\theta n_i}+h.c.) +\sum_{i}h_i.
\label{BH}
\ee
\begin{figure}[t]
\includegraphics[width=7cm, height= 4.5 cm]{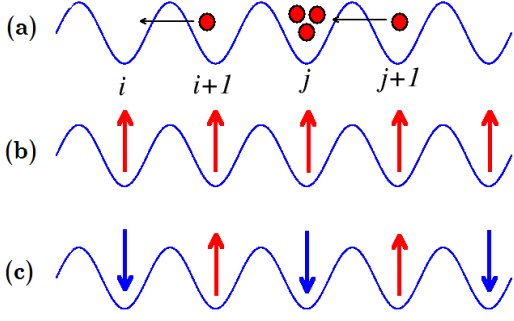}
\caption{(a) The illustration of the conditional effect of $e^{\theta n_i}$,
$n_i=0$,  $e^{\theta n_i}=1$; $n_j=3$, $e^{\theta n_j}\ne1$.
(b) The arrows with same length and directions represent the
distribution of expectation values of $b_i$ in the homogenous SF phase.
Generally, $\langle b_i\rangle$ is a complex number. (b) In the BSF phase, the arrows with same length but opposite directions means that
the distribution of expectation value of $b_i$ is in a staggered pattern.}
\label{con}
\end{figure}
Fig.~\ref{con} (a) shows the conditional effects of the  density-dependent phase factor.
The effects of the phase are caused by $b^{\dagger}_{i}b_{i+1}e^{i\theta n_i}$.
If there are no particles in the site $i$, namely $n_i=0$, then the phase factor is still given by $e^{i\theta n_i}=1$.
In this way, the model is no different from the Bose-Hubbard model.

The situation becomes different for a soft-core Bose-Hubbard model.
If three particles already exist in the site $j$, the phase factor becomes
$e^{i\theta n_j}=e^{i3\theta}$.

Figs.~\ref{con} (b) and (c) show the typical effects of $e^{i\theta n_i}$.
In a homogenous SF phase,
the $\langle b_i\rangle$ are   distributed homogenously and represented
by arrows with the same direction (imaginary and real parts) and length (value), and $\langle b_i\rangle$ generally is a complex number.

For some values of $\theta\ne 0$, there is a translational broken symmetry of
the distribution of the expectation value for $b_i$, characterized by an oscillating sign but with the same  values, i.e., $\langle b_i\rangle=-\langle b_{i+1}\rangle$. The SF phase with this property is called a broken-symmetry superfluid (BSF).

\subsection{MF and DMRG methods}
\label{sec:method}
According to previous studies\cite{sta}, in order to get the Hamiltonian in the MF frame and
for convenience, the following term\cite{sta}
$ b_j^{\dagger}e^{i\theta n_j}b_{j+1}= c_{j}^{\dagger}b_{j+1}
$
is defined and decoupled as
\[ c_{j}^{\dagger}b_{j+1}\approx-\Psi_{2,j}^*\Psi_{1,j+1}+\Psi_{2,j}^{*}b_{j+1}+c_j^{\dagger}\Psi _{1,j+1}~,
\]
where the order parameters are $\Psi_{1,j}=\la b_j\ra$ and $ \Psi_{2,j}=\la c_j\ra $.
Without the nearest repulsion, the system looks homogenous  and accordingly, the Hamiltonian of Eq.~(\ref{BH}) in the MF frame becomes
$
H=\sum_{j}H_j
$
with
\be
H_j=h^s-t(\Psi_2b^{\dagger}+\Psi_2^{*}b+\Psi_1c^{\dagger}+\Psi_1^{*}c-\Psi_{1}^*\Psi_{2}-\Psi_2^{*}\Psi_1).
\label{hs}
\ee
In the equation above, Ref.~\cite{sta} neglects the subscript $j$ as the order parameters are homogenous, i.e:  $\Psi_{1}=\la b_j\ra=\la b_{j+1}\ra$,   $\Psi_2=\la c_j \ra=\la c_{j+1} \ra $.
However, this artificial homogenous condition is too strong to account for some interesting nonuniform phases.
Therefore, it is necessary to use subscripts $A$ and $B$ to distinguish
the physical quantities $\Psi_{1}$ and $\Psi_{2}$, on the different sublattices, such as
$\Psi_{1A}$, $\Psi_{1B}$,  $\Psi_{2A}$, and $\Psi_{2B}$.

In the MF frame, we assume only a
two-sublattice structure as a possible inhomogeneity in the ground
state.
 However, due to the existence of the phase factor $\theta$ of
model (\ref{BH}), it is naturally expected to obtain a state with longer
structures. This strong constraint is overcome by the DMRG method.

We define the average density of atoms on both sublattices as
$\rho_{A}=\langle n_A \rangle$ and $\rho_B =\langle n_B\rangle$.
Combining Eq.~(\ref{hs}) and the definitions of order parameters, we obtain the local Hamiltonian on the sublattice A and thus
\begin{equation}
\begin{aligned}
H_A&=-\frac{zt}{2}[c^{\dagger}_{A}\Psi_{1B}+c_{A}\Psi^{*}_{1B}+b_{A}\Psi^{*}_{2B}+b^{\dagger}_{A}\Psi_{2B}\\
&-\frac{1}{2}(\Psi^{*}_{2A}\Psi_{1B}+\Psi_{2A}\Psi^{*}_{1B}+\Psi^{*}_{2B}\Psi_{1A}+\Psi_{2B}\Psi^{*}_{1A})]\\
&+\frac{U}{2} n_{A}(n_A-1)-\mu n_{A},
\label{ha}
 \end{aligned}
\end{equation}
and the Hamiltonian on $H_B$ is
\begin{equation}
\begin{aligned}
H_B&=-\frac{zt}{2}[b^{\dagger}_{B}\Psi_{2A}+b_{B}\Psi^{*}_{2A}+c^{\dagger}_{B}\Psi_{1A}+c_{B}\Psi^{*}_{1A}\\
&-\frac{1}{2}(\Psi^{*}_{2A}\Psi_{1B}+\Psi_{2A}\Psi^{*}_{1B}+\Psi^{*}_{2B}\Psi_{1A}+\Psi_{2B}\Psi^{*}_{1A})]\\
&+\frac{U}{2} n_{B}(n_{B}-1)-\mu n_{B}
\label{hb}
\end{aligned}
\end{equation}
By solving eqs.~(\ref{ha}) and (\ref{hb})
self-consistently,
we reproduced the consistent phase dia\-gram of Ref.~\cite{sta}, which is not shown here.

To confirm the results obtained by the MF method, we also use
the DMRG method.
To deal with the complex tunneling element, we combine
$e^{i\theta n_{j}}$ and $b_j^{\dagger}$ into one operator.
If $\theta\ne 0$, the Hamiltonian becomes complex but remains Hermitian nonetheless.
We just use a rapid prototyping program like MATLAB to get the ground-state energy and wave function.
The periodic boundary condition is used to suppress the boundary effects.

\subsection{The sampled quantities}

In the MF method, the particle density is $\rho=(\rho_A+\rho_B)/2$ and the SF density is $\Psi=|\Psi_{1A}+\Psi_{1B}|/2$.
In this model, there are two types of superfluid phase: the SF phase and the BSF phase.
With the MF method, the SF phase is characterized by  $\Psi \ne 0$ and the BSF phase is denoted by $\Delta\Psi=|\Psi_{1A}-\Psi_{1B}|\ne 0 $.

With the DMRG method, the correlation $C(r)=\langle b_i^{\dagger}b_{i+r}\rangle$ and the average correlation
$C=\sum_{r=1}^{L}C(r)/L$ are calculated. The momentum distribution is defined as
$
n(k)=\frac{1}{L}\sum_{i,j}\langle b_i^{\dagger}b_{j}\rangle e^{ik(i-j)}
$\cite{momentum}.

\section{mean field results}

\label{sec:result}

\subsection{Staggered distribution of the SF and BSF phases}

\begin{figure}[htb]
\includegraphics[width=0.35\textwidth, angle=270]{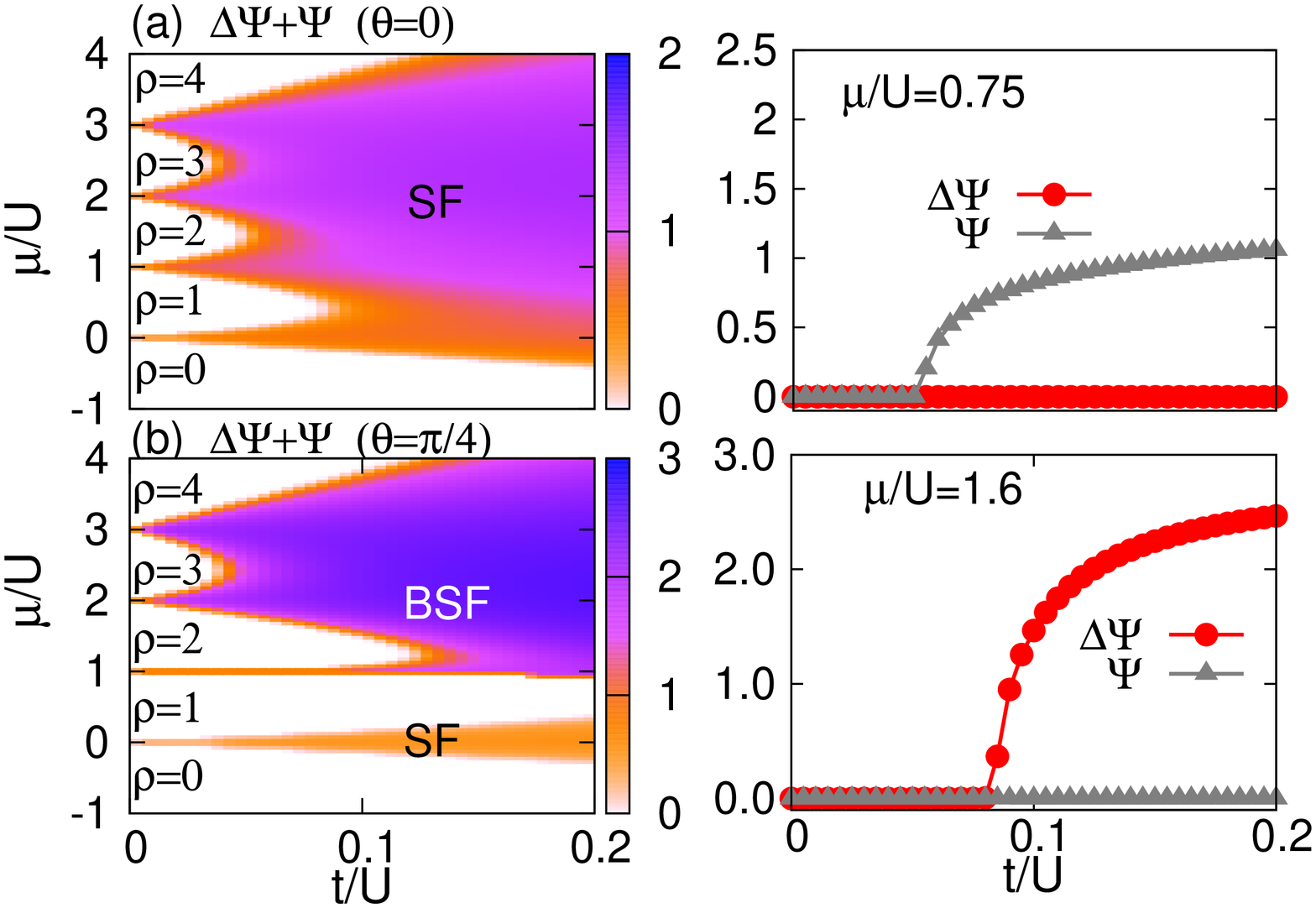}
\includegraphics[width=0.35\textwidth, angle=270]{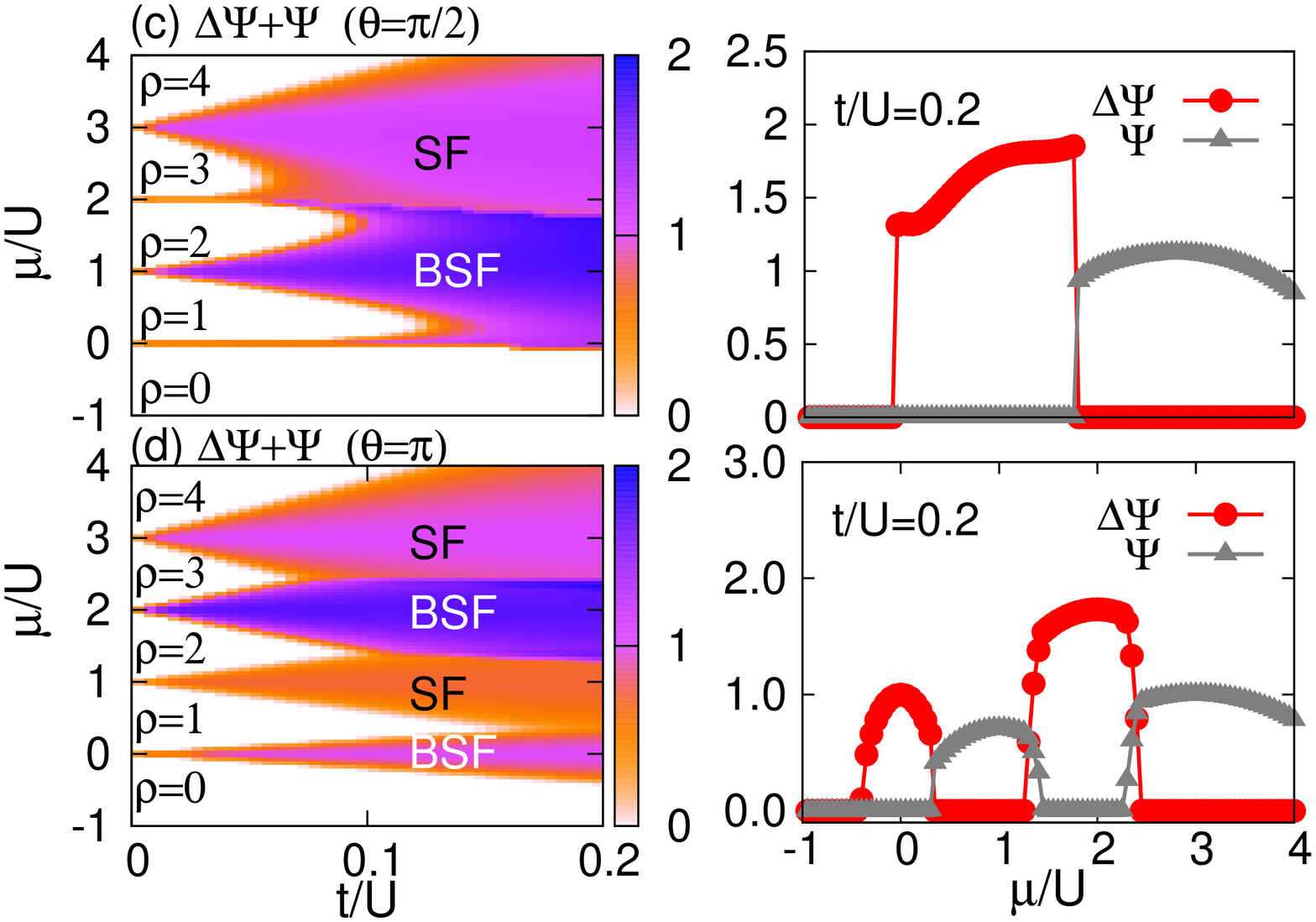}
\caption{(Color online) The quantum phase ( $\Psi+\Delta\Psi$), which contains the SF, BSF and MI phases in the plane ($t/U$, $\mu/U$) of the model with  (a) $\theta =$ 0, (b) $\pi/4$, (c) $\pi/2$, and (d) $\pi $,  on the left column, from top to bottom sequentially. The right column are detailed descriptions of $\Psi$  and $\Delta \Psi$ along $t/U$ or $\mu/U$.
}\label{v0mf}
\end{figure}

In this section, we  present  the global phase diagrams, by plotting
 $\Psi+\Delta \Psi$ in the plane  ($t/U$, $\mu/U$) for $\theta$ at $0$, $\pi/4$,
$\pi/2$ and  $\pi$. The phase diagrams contain the SF and BSF phases and the staggered distribution between both phases.
Fig.~\ref{v0mf} (left) shows the phase diagrams of the model, the right column shows the detailed descriptions of $\Psi$ and $\Delta\Psi$ along $t/U$ or $\mu/U$.

At small $t/U$,  with  the maximum  on-site occupation being $n_{max}=4$, the MI phases emerge
sequentially  with
densities $\rho=0, 1, 2, 3$ and $4$ when the chemical potential $\mu/U$ increases from $0$ to $4$.

At larger $t/U$, the system sits in the SF phase or the BSF phase, which are labeled in the phase diagrams.
In the work of Ref.~\cite{sta}, the SF-MI phase transition boundaries have been obtained with $t/U>0$ and different values of $\theta$.
The boundary lines are consistent with the results of Ref.~\cite{sta}, which are not shown here.

In Fig.~\ref{v0mf}(a), for $\theta=0$, the SF phase emerges with finite values of $t/U$.
As shown in the right column
at $\mu/U=0.75$,
in the range $t/U>0.05$ , the SF phase is  localized
with $\Psi\ne0$ and $\Delta\Psi=0$.
By increasing $t/U$, $\Psi$
changes continuously into a non-zero regime, which means the MI-SF phase
transition is continuous.

For $\theta=\pi/4$, when compared with $\theta=0$, a staggered pattern emerges for the distribution between the
BSF and SF phases. The BSF phase emerges in the top right part of the phase diagram,  and the SF phase emerges in the lower  part, respectively.
At the same time, the BSF and SF phases are separated by the MI phases.

For $\theta=\pi/2$, the BSF and SF phases can join together.
To show the details, we scan $\mu/U$ along a cut line $t/U=0.2$.
In the region $-1\textless\mu/U\textless-0.2$, both quantities $\Psi$ and $\Delta\Psi$ are equal to zero.
By increasing $\mu/U$, the quantity $\Delta\Psi$ becomes nonzero with an obvious jump.
This jump is due to the finite size of the MF frame, and finally disappears according to the DMRG calculation (not shown). The phase transition is still continuous.

In Fig.~\ref{v0mf}(d), for $\theta=\pi$,
 more of the SF and BSF phases emerge from top to bottom.
The phase transition between the BSF and SF phases is first order because of the jumps within $\Psi$ and $\Delta\Psi$.

\begin{figure}[t]
\includegraphics[width=0.45 \textwidth]{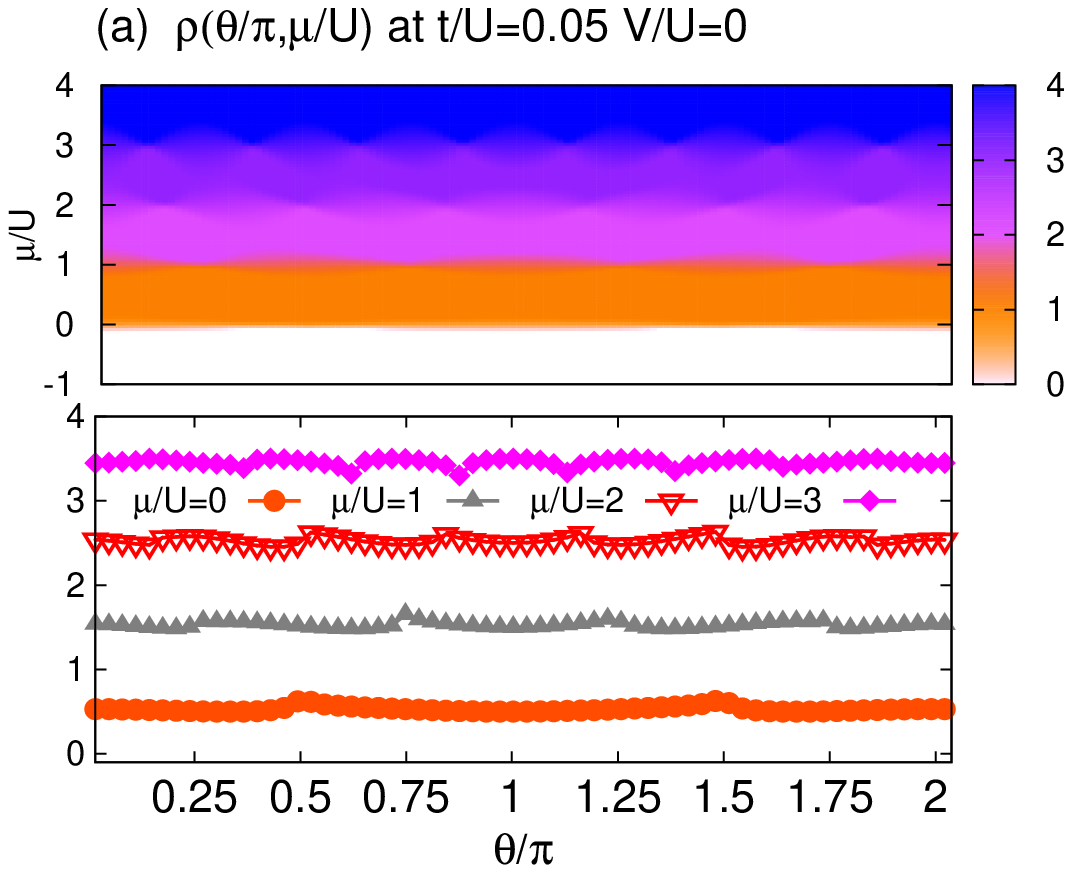}
\includegraphics[width=0.45 \textwidth]{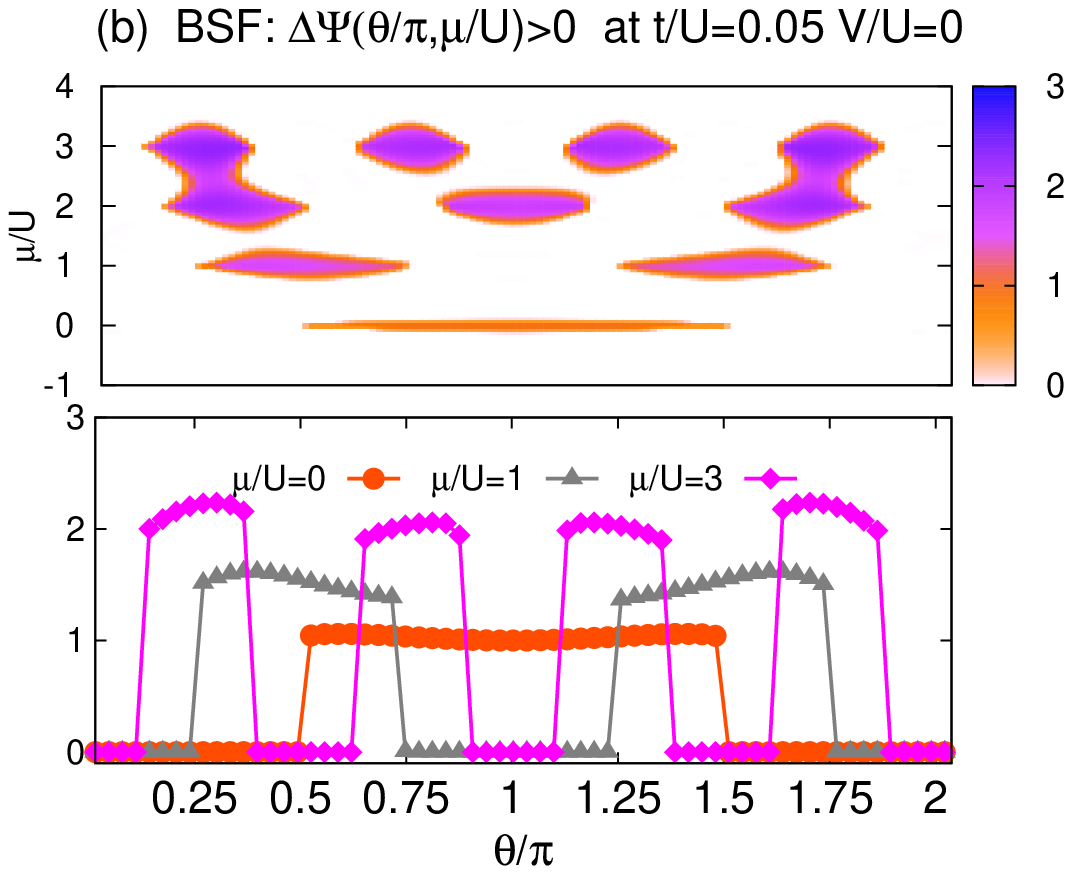}
\caption{(Color online) The values of (a) density $\rho$ and (b) $\Delta \Psi$  in the plane ($\theta/\pi$, $\mu/U$) at $t/U=0.05$ and $V/U=0$.
At the bottom of each figure (in color), we also plot the three quantities at $\mu/U=0$, 1, and 3 as a function of  $\theta/\pi$ .
}\label{v0rhopsi}
\end{figure}
\subsection{ $t/U=0.05$: first-order BSF-SF phase transition}

In the section above, we just show the quantities with four discrete values of $\theta$.
Continuous  modulation of $\theta$,
Figs.~\ref{v0rhopsi} (a)-(b), show the quantities $\rho$ and $\Delta\Psi$ (from top to bottom,
respectively) with the parameter plane ($\theta/\pi$, $\mu/U$) at
$t/U=0.05$.

In Fig.~\ref{v0rhopsi} (a), the distribution of $\rho$ is shown as a function of
$\theta/\pi$ and $\mu/U$.
By increasing $\mu/U$ sequentially from 0 to 4, in the direction of $\mu/U$,
the color becomes darker and darker, which means the density
grows. We also find the ``wave" along the boundaries between different colors (densities) along the $\theta/\pi$ direction.
However, as we scan $\theta/\pi$ at the bottom of Fig.~\ref{v0rhopsi} (a), the densities  have several kinks.
The emergence frequency of the kinks increases
as the density $\rho$ (chemical potential $\mu$) increases.
The number of kinks for  $\mu/U=0$, 1, 2 and 3  are
2, 4, 6 and 8, respectively.

For example, for $\mu/U=0$ by changing $\theta/\pi$,
the density curve has two kinks at $\theta/\pi=0.5$ and 1.5.
Actually the kinks emerge as
a phase transition takes place.
For larger $\mu/U$, the variation becomes
more obvious.
This phenomena can be understood by the following.
According to $a_j=b_jexp(i\theta\sum_i^{j-1} n_i)$,
if the value of $\sum_i^{j-1} n_i$  is bigger, then
the operator $a_j=b_jexp(i\theta\sum_i^{j-1} n_i)$ will change
more quickly as $\theta$ changes.

In Fig.~\ref{v0rhopsi} (b), the distribution of $\Delta\Psi$ is shown in
the plane ($\mu/U$, $\theta/\pi$). All colored regions ($\Delta\Psi$) represent the BSF phase.
For example, when $\mu/U=0$, $\Delta\Psi>0$ in a narrow area in the regime $0.5<\theta / \pi < 1.5$.
We also show $\Delta\Psi$ at $\mu/U=0, 1$ and 3 along $\theta/\pi$, which
confirms the result from Fig.~\ref{v0rhopsi} (a).
An obvious first-order SF-BSF phase transition is found  because of
the jumps of the order parameters.
The distribution of $\Psi$ and the details are not shown here.

\section{DMRG Results}
\label{sec:dmrg}
\subsection{Phase diagrams}

\begin{figure}[b]
\includegraphics[width=0.45 \textwidth, angle=0]{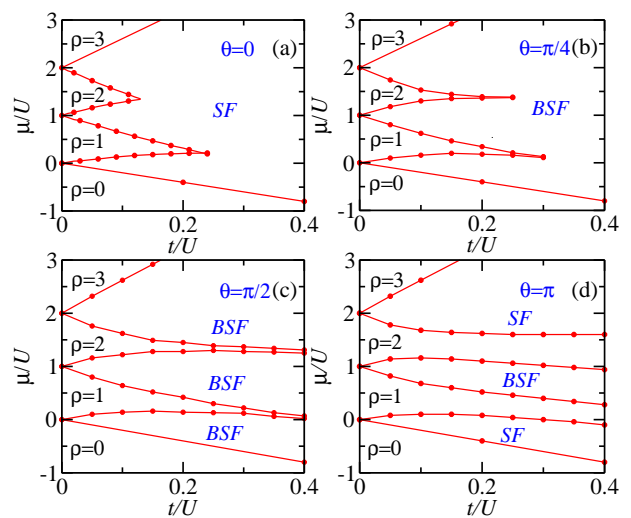}
\caption{(Color online) The DMRG phase diagrams, which contains the SF, BSF and MI phases in the plane ($t/U$, $\mu/U$) of the model with (a) $\theta =0$, (b) $\pi/4$, (c) $\pi/2$ and (d) $\pi $. }
\label{v0dmrg}
\end{figure}
\begin{figure}[t]
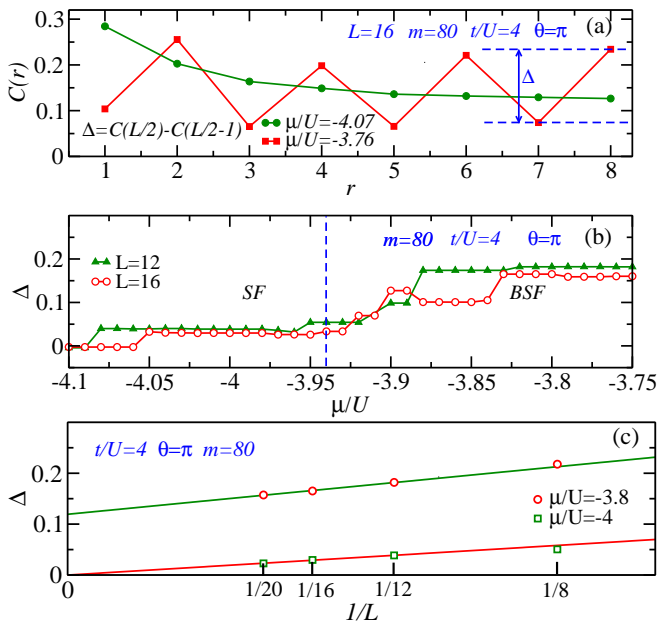

\includegraphics[width=1\columnwidth, angle=0]{fig5a.eps}
\includegraphics[width=1\columnwidth, angle=0]{fig5b.eps}
\caption{(Color online) (a) Correlation C(r) with system size $L=16$,  $t/U=4$, $\theta=\pi$  in the SF phase at $\mu/U=-4.07$ and in the BSF phase with $\mu/U=-3.76$,  (b)  $\Delta$ .vs. $\mu/U$  with system sizes $L=12,  16$. }
\label{sfbsf}
\end{figure}

Fig. \ref{v0dmrg} shows the phase diagrams, which contain the SF, BSF and MI phases in the plane ($t/U$, $\mu/U$) of the model with $\theta =0, \pi/4, \pi/2$, and $\pi $.

For $\theta =0$, only the SF phase emerges. This is very consistent
with the corresponding result in the MF frame, as shown in Fig. \ref{v0mf} (a).

For $\theta =\pi/4$ and $\pi/2$, the MF phase diagrams contain both the SF and BSF phases. However, compared with Fig.~\ref{v0mf}, the DMRG only detects the BSF phases. The SF phase detected by MF method is actually the BSF phase, where the wavelengths
of the waves  emerging in the correlation are too long to be detected by the MF method.

For $\theta=\pi$, the stagger distributions of the SF and BSF phases  are also found by the DMRG calculation.

With larger $t$ (for example $t/U=4$), the direct SF-BSF phase transitions occur.
In Fig.~\ref{sfbsf} (a), two typical correlations of both the SF and BSF phases are shown.
In the BSF phase, the correlation exhibits  obvious oscillations, which can be roughly characterized by $\Delta=C(L/2)-C(L/2-1)$.
Clearly, $\Delta \ne 0$ emerges in the BSF phase and $\Delta = 0$ emerges in the SF phase. In Fig.~\ref{sfbsf} (b),  we find $\Delta = 0$
if $\mu/U\textless -3.94$ while $\Delta \ne 0$  when $\mu/U> -3.94$ in the thermodynamic limit. The finite size scaling analysis is performed
at  $\mu/U=-4$ and $-3.8$ as shown in Fig.~\ref{sfbsf}(c).  Although there is no rough jump of the order parameter $\Delta$, in contradistinction to the MF prediction,  one can still observe a direct phase transition between the two type of superfluid phases.

\subsection{Beat and Correlation}
To clearly see the effects of $\theta\ne0$, we choose $\theta=\pi/2$. In this case, more interesting properties emerge.
Firstly,  no homogenous SF phase exists. Except for the MI phases
with different fillings, all regimes are in the BSF phase. This is obviously different from the MF result in Fig.~\ref{v0mf} (b).

\begin{figure}[t]
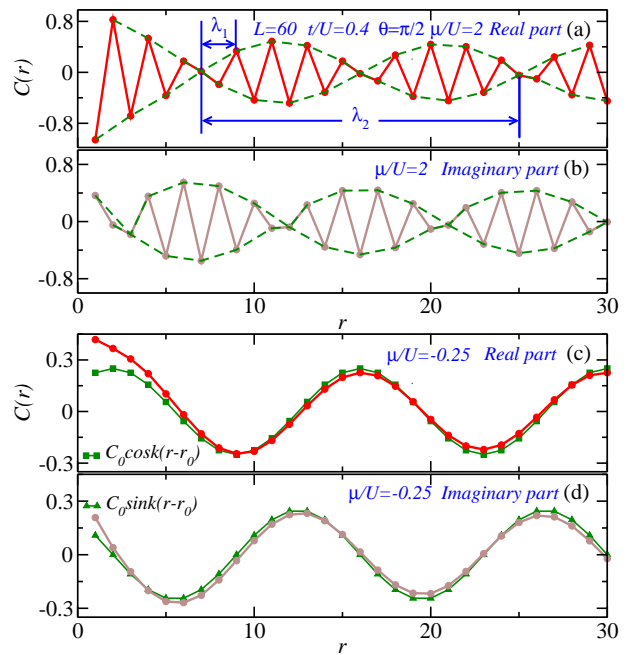

\includegraphics[width=0.45 \textwidth]{fig6a.eps}\hskip 0.5cm
\includegraphics[width=0.45 \textwidth]{fig6b.eps}\hskip 0.5cm

\caption{(Color online) Correlations $C(r)$ for $L=60$, $m=80$ $\theta=\pi/2$, $t/U=0.4$. (a) Real part of $C(r)$ with $\mu/U=2$ (b) Imaginary part $C(r)$
with $\mu/U=2$ (c) Real part of $C(r)$ with $\mu/U=-0.25$ (d) Imaginary part $C(r)$
with $\mu/U=-0.25$. In (a), $\lambda_1$ and $\lambda_2$ are  the  oscillation wavelength and beat wavelength respectively. The green lines are plotted to emphasize the beats. }
\label{co2}
\end{figure}

The properties of the BSF phase are studied by plotting the
correlation along $t/U= 0.4$ from $\mu/U=-0.5$ to $\mu/U=3$ at some intervals, is chosen to show in Fig.~\ref{co2}.
The first finding is the beat phenomena emerging from the correlation.
Furthermore, the  oscillation period of the correlation  becomes longer or shorter  as
 the density (chemical potential) changes. Moreover, the type of  behavior of the correlation emerges in a staggered pattern in the phase diagram.

The superposition of two waves of the same frequency propagating in opposite directions will cause a standing wave, in which the maximum amplitude and minimum amplitude are constants.
If the two waves have slightly different frequencies, beats will forms, in which the maximum and minimum amplitudes are no longer constants.

Fig.~\ref{co2} (a) shows the real part of the correlation $C(r)$ at $\mu/U=-0.25$ and $t/U=0.4$ with size $L=60$.
Clearly, the sign of the correlation oscillates as the distance $r$ grows between the two sites. It behaves in a triangular wave shape with a beat.
The correlation increases once and decreases once, backwards and forwards, where the oscillation wavelength is $\lambda_1=2$, and the beat wavelength is $\lambda_2= 18$. In the position $r=16$, the amplitude tends to zero, where zero is the node of a beat.
Beyond the node, the amplitude grows again and a  new  beat starts again. Even at the ends, beats are discernable, because the nodes and maximum are observable.

Fig.~\ref{co2} (b) shows the imaginary part of  the correlation $C(r)$. The positions of the nodes in the real part of $C(r)$ correspond to the peaks or the lowest positions in the imaginary part.
 In the mean-field frame, we assume only a two-sublattice structure as a possible inhomogeneity in the ground state. However, due to the existence of the phase factor $\theta$ of model (\ref{BH}), it is naturally expected that a state with a longer
(incommensurate) wavelength can appear. For example, the order
parameters may have uniform amplitude but with a ``spiral" phase factor, i.e.,
$\Psi \propto |\Psi|e^{ ik r}$.  The DMRG method can overcome the constraint from the MF method. From  our calculation, $C(r)$ realy emerges according to a pattern of  \be C(r) = C_0e^{i k (r-r_0)}. \ee For example, in the
case of the green lines of Figs.~\ref{co2}(c) and (d),  $C_0=0.25$ and $k=0.4488$ are used. Here, $r_0=2$ and $r_0=9$ are
for the real and imaginary parts, respectively.  Furthermore, in Figs.~\ref{co2} (a) and (b), the correlation $C(r)$ in the shape of the beats obey the equation as follows  \be C(r) = C_0e^{i k (r-r_0)}(-1)^{r}. \ee

\noindent
For  $\mu/U=1.5$, the wavelengths of the beats $\lambda_2$ become shorter
$\lambda_2=10$.
From $\mu/U=0.5$ to $-0.5$, $\lambda_2$  disappears and therefore beats don't exist. The value of the oscillation  wavelength $\lambda_1$ is still present.

\begin{figure}[t]
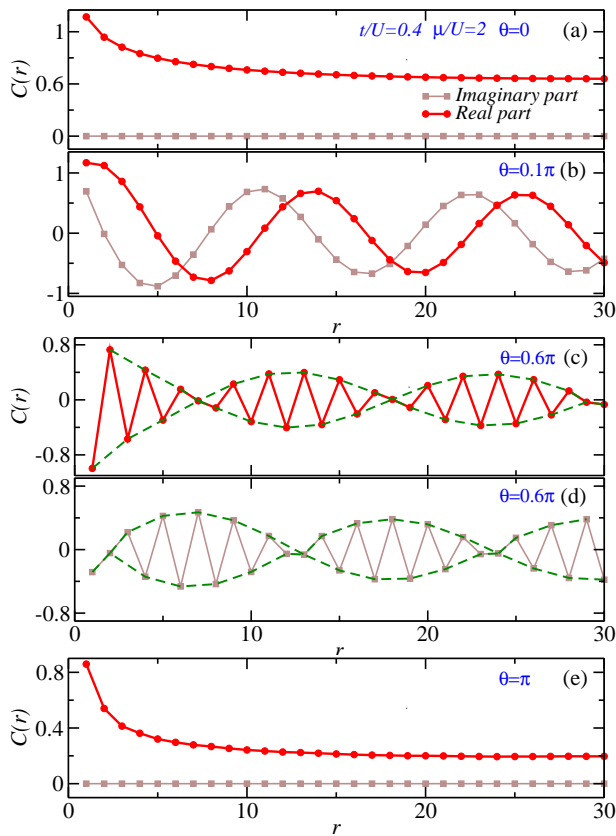

\includegraphics[width=0.45 \textwidth]{fig7ab.eps}\hskip 0.5cm
\includegraphics[width=0.45 \textwidth]{fig7cd.eps}\vskip -0.05cm
\includegraphics[width=0.45 \textwidth]{fig7e.eps}\hskip 0.05cm
\caption{(Color online) Emergence and disappearance of beats from both of the real and imaginary parts of the correlation $C(r)$ by modulation of $\theta$ for $\theta/\pi=0$, $0.1$, $0.6$ and $1$ at $t/U= 0.4$, $\mu/U=2$.}
\label{thetadmrg}
\end{figure}

Apart from the notion that $\mu/U$ or the density will change the properties with or without beats, the effects of $\theta$ upon the correlation
need to be discussed. In Figs.~\ref{thetadmrg} (a)-(d), we start with a SF phase, at $t/U=0.4$, $\mu/U=2$, and $\theta=0$,  no oscillation in the correlation exists.
We then increase $\theta/\pi$ from $0$ to $1$ with a spacing of  $0.1$. When $\theta=0.1 \pi$, the correlation oscillates smoothly.
When $\theta=0.6\pi$, beats emerge. Beats disappear at $\theta=\pi$.

To summarise, for $0\textless\theta\textless\pi$, the BSF phase emerges and beats emerges for a range of values of $\mu/U$.

\subsection{Explanation  of  beats by the momentum distribution}
\begin{figure}[tb]
\includegraphics[width=0.45 \textwidth]{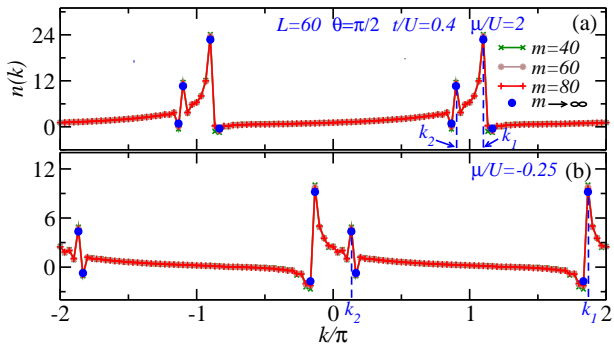}\hskip 0.5cm
\caption{(Color online) Asymmetric momentum distribution $n(k)$ by DMRG calculations with the same parameters in Figs.~\ref{co2}. (a)$k_1=1.1$ and $k_2=0.9$, satisfying the existing of
beats $y<1/3$; (b) $k_1=1.87$ and $k_2=0.13$, which does not satisfy $y<1/3$.
}
\label{mdmrg}
\end{figure}

The reason why beats exists in Fig.~\ref{co2} (a) or do not exist in Fig.~\ref{co2} (b)
can be analyzed by the momentum distributions.
 Figs.~\ref{mdmrg} (a) and (b) show the momentum distributions $n(k)$ with the same parameters as those of Figs.~\ref{co2}, which are
helpful to us in understanding the behavior of the correlations.
On the whole, the two peaks of the momentum distribution reflect the wave numbers ${k}_1$ and ${k}_2$, which superpose together
to form various kinds of correlation patterns. The condition for beat existence is given by
 $y=\frac{{k}_1-{k}_2}{{k}_1+{k}_2}<\frac{1}{3}$ (see appendix).

To check the correction of the momentum distribution obtained, we sum $n(k)$ over different values of the wave numbers ${k}$ as follows
\be
\sum_{m=1}^{L} n(k=\dfrac{2m\pi}{L})=L\rho=N,
\ee
where $L$ is the chain length and $N$ is the number of total particles. Our numerical values are very consistent
with the above equation.

\begin{table}[b]
 \caption{
 Values of $\lambda_1$, $k_1^{'}/\pi$, $k_1/\pi$, $\lambda_2$, $k_2^{'}/\pi$, and $k_2/\pi$ in Fig.~\ref{mdmrg}.
 $\lambda_1$ and  $\lambda_2$ are the oscillation  and beat wavelengths, respectively. $k_1$ and $k_2$ are from the peaks
 of momentum distributions. $k_1^{'}$ and $k_2^{'}$ are from the real space correlation. 
 }
 \begin{center}
\begin{tabular}{rccccccc}
  \hline
  \hline
 $\mu/U$&~ $\lambda_1$ ~&~ $k_1^{'}/\pi$ ~&~ ~$k_1/\pi$~&~ $\lambda_2$ ~& ~$k_2^{'}/\pi$~ ~&~ ~$k_2/\pi$~ ~&~ ~$y$~ \\ \hline
1.5 &2& 1.20  &1.17&10 &0.80 &0.83  &0.17\\
2.0 &2& 1.11  &1.10&18 &0.89 &0.90  &0.1 \\
 2.5&2& 1.06  &1.03&34 &0.94 &0.97  &0.03\\
$-0.5$ & &      &1.97&  &    &0.03  &0.97\\
$-0.25$ & &      &1.87&  &    &0.13  &0.87\\
0.0 & &      &1.73&  &    &0.27  &0.73\\
0.5 & &      &1.43&  &    &0.57  &0.43\\

  \hline
  \hline
\end{tabular}
\label{Tab:t1}
\end{center}
\end{table}

In Fig.~\ref{mdmrg} (a), for $\mu/U=2$, the peaks $k_{1}$ and $k_{2}$ can be obtained from
the momentum distribution as shown in table \ref{Tab:t1}, where $k_1=1.1$ and $k_2=0.9$.
Beats form clearly because $y=0.1\textless1/3$.
 To check the correctness of $k_{1}$ and $k_{2}$,
the peaks can also be compared with $k_{1}^{'}$ and $k_{2}^{'}$ by deviation from the oscillation lengths $\lambda_1$ and
the beat length $\lambda_2$ in real space. The flow chart is as follow,
\be
\lambda_1,\lambda_2\Rightarrow k_{quick},k_{slow}\Rightarrow k_1^{'},k_2^{'}\Longleftrightarrow k_1,k_2.
\ee
We assume a beat,  resulting from two superposed waves with slightly different frequencies $k_1^{'}$ and $k_2^{'}$, then
we will obtain a beat with an oscillation  frequency  $k_{quick}=\frac{k_1^{'}+k_2^{'}}{2}$ and a beat frequency $k_{slow}=\frac{k_1^{'}-k_2^{'}}{2}$, where $k_{quick}$ and $k_{slow}$ can be obtained by
counting $\lambda_1$ and $\lambda_2$, where
$k_{quick}=\frac{2\pi}{\lambda_1}$ and
$k_{slow}=\frac{2\pi}{\lambda_2}$.

Apparently, as shown in table \ref{Tab:t1}, by comparing  each of the wave numbers ${k}_i$ ($i=1,2$) of the same index in table \ref{Tab:t1}, we find
that $k_1$ and $k_1'$, and $k_2$ and $k_2'$
are fairly close to each other to within the first two digits.
Two ways of obtaining the frequencies of the two superposed waves are checked against each other.
The finite size effects and quantum fluctuation make ${k}_i$ and ${k}_i'$ ($i=1,2$) a little different.

Fig.~\ref{mdmrg} (b) shows that, at $\mu/U=-0.25$, two separated peaks far apart from each other emerge around $k_2/\pi=0$ and $k_1/\pi=2$,
where the wave numbers  ${k}_1$ and ${k}_2$ are available in Table \ref{Tab:t1}.
In real space, the beat phenomena does not exist as shown in Fig.~\ref{co2} (b) with the same parameters.
The reason is not that the length of the beat is too long to be seen in a limited size $L=60$, where  $L$ is supposed to be the length of the system.  Rather it is because the two wave numbers ${k}_1$ and ${k}_2$ do not satisfy the condition for existence of the beat.

\subsection{Asymmetry of momentum distribution}

In table \ref{Tab:t1}, numerical results mean that   the sums over  ${k}_1$ and ${k}_2$ remain at $2\pi$ and
 the two wave numbers are symmetric with $k/\pi=1$. However, the shapes of the momentum distributions are asymmetric with $k=0$.
The reflectional symmetry about $k=0$ is broken because of
the asymmetric phase factor assigned to the hopping.
The asymmetry is consistent with the results in Refs\cite{ground1,ground2,tang}.

To understand the asymmetry of the momentum distribution, the asymmetry of
the energy spectrum in momentum space is given.
It is well known that, when $\theta=0$,  the energy of the
non interacting Bose-Hubbard model with Hamiltonian $H^b =-t\sum_{i=1}^{L}(b^{\dagger}_{i}b_{i+1}e^{i\theta n_i}+h.c.)$ is $E(k)=-2t \cos (k)$, which is obviously
symmetric about $k/\pi=0$.
In the derivation, the relationship
\be
\sum_i(b^{\dagger}_{i}b_{i+1}+b^{\dagger}_{i}b_{i-1})=\sum_kb^{\dagger}_{k}b_{k}(e^{ik}+e^{-ik}) \label{eq:11}
\ee
is used\cite{solidth}.
For a system with a fixed density, $n_i$ is a constant $\overline n_i$.  Letting $e^{i\theta \overline n_i}$ couple the eq.~(\ref{eq:11}), we get
\be
\begin{aligned}
&\sum_i(b^{\dagger}_{i}b_{i+1}e^{i\theta \overline n_i}+b^{\dagger}_{i}b_{i-1}e^{-i\theta \overline n_i})\\
&=\sum_kb^{\dagger}_{k}b_{k}(e^{ik+i\overline n_i\theta}+e^{-ik-i\overline n_i\theta})
\end{aligned}
\ee
Therefore, $E(k)=-2t \cos (k+\theta \, \overline n_i)$ and should be asymmetric with $k=0$ if $\theta \, \overline n_i\ne 0$ .
\section{Discussion and conclusion}

\label{sec:con}

By using the DMRG and MF methods,
the anyon Hubbard model has been studied systematically on a one dimensional lattice.

The MF method can provide us with the basic phase diagrams, which are consistent
with the results from the DMRG method with $\theta/\pi=0$.
For other values of $\theta$, although the MF method cannot provide the precise phase-diagrams, the MF method
still help us search for the different  behaviors of the  correlations.

The concept of broken-symmetry plays an important role in  theoretical physics, such as in the origin of the mass associated with the Higgs boson\cite{bs1}.
Here, various interesting patterns of the correlation $b^+_ib_{i+r}$  enrich  the concept of broken-symmetry in correlated boson systems.
In some areas, the correlation yields beats if the two supposing wave numbers ${k}_1$ and ${k}_2$ satisfy
$({k}_1-{k}_2)/({k}_1+{k}_2)<1/3$.

We never see beats in the correlation $b^{\dagger}_ib_{i+r}$ for the usual Bose-Hubbard model except the solid order pattern\cite{soli}.
Note that this work is the first to  observe beats of the correlation  $b^{\dagger}_ib_{i+r}$  in the Bose-Hubbard type model.
Different kinds of momentum distributions are analysed  and  expected to be observed in optical lattice experiments.

\acknowledgments
We thank Sebastian Greschner and Guixin Tang
for their invaluable discussions as well as their correlation data for comparison.
We also thank  Min Gong for his helpful suggestions.
W. Zhang  is supported  by the NSFC under Grant
No.11305113,  Youth Foundation of Taiyuan University of Technology 1205-04020102.
T.C. Scott is supported in China by the project GDW201400042 for the ``high end foreign experts project''.
 Y.  Zhang is supported by NSF of China under Grant Nos. 11234008 and 11474189, the National Basic Research Program of China (973 Program) under Grant No. 2011CB921601, Program for Changjiang Scholars and Innovative Research Team in University (PCSIRT)(No. IRT13076).

\appendix
\section{the standard $y<1/3$}
Here, we show how to get the criteria of existence of a beat, namely, $y< 1/3$.
We assume a beat mixed with two  waves with wave numbers  ${k}_1^{'}$ and ${k}_2^{'}$, respectively.
The difference of the two wave numbers should be less than the sum of both wave numbers, namely
\be
k_1^{'}-k_2^{'}<k_1^{'}+k_2^{'}.
\label{1}
\ee

For convenience, we let   $y=\frac{k_1^{'}-k_2^{'}}{k_1^{'}+k_2{'}}<1$, and  $k_1{'}=ak_2^{'}$. Then we  assume  $\dfrac{k_1{'}}{k_2^{'}}=a>1$, which
leads to
\be
y=\dfrac{ak_2^{'}-k_2^{'}}{ak_2^{'}+k_2{'}}=\dfrac{a-1}{a+1}< 1
\label{2}
\ee
Now, we discuss the possible value of $a$.
Firstly, a beat will not exist if the two wave numbers are the same, ie., $a=1$ or
one of the wave numbers ${k}_1^{'}$ is twice as much as that of the other wave numbers ${k}_2^{'}$, namely, $a=2$.
A reasonable choice of $a$ is  $1<a<2$ and then we can easily obtain $0<y<\dfrac{1}{3}$ \cite{beat}.
\section{Comparison with Greschner Data}
To check correctness of our findings, we compare the data of $C(r)$  with the same boundary conditions (periodic boundary conditions) from
the data  of Greschner, an author of Ref.~\cite{santos2}. The parameters are $L=60$, $U=2.5$, $t=1$ and $n_{max}=4$.   Both data have beats and are basically consistent with each other quantitatively although we used $\mu/U=3.3455$ and the number of the particle is $N_{total}=137$ and S. Greschner used $N_{total}=139$.

\begin{figure}[tb]
\includegraphics[width=0.45 \textwidth]{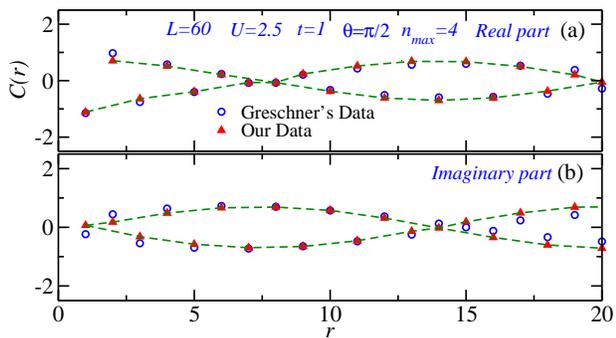}
\caption{(Color online) Comparison of our data $C(r)$ with the result from S. Greschner\cite{santos2}.}
\label{com}
\end{figure}

\end{document}